\documentclass[twocolumn,showpacs,preprintnumbers,amsmath,amssymb]{revtex4}
\usepackage {longtable}
\usepackage{graphicx}
\usepackage{dcolumn}
\usepackage{bm}

\begin{document}

\title{Quasiclassical calculations of BBR-induced depopulation
rates and effective lifetimes of Rydberg \textit{nS}, \textit{nP},
and \textit{nD} alkali-metal atoms with $n \le 80$.}

{\author{I.~I.~Beterov}
\email{beterov@isp.nsc.ru}

\author{I.~I.~Ryabtsev}

\author{D.~B.~Tretyakov}

\author{V.~M.~Entin}

\affiliation{Institute of Semiconductor Physics, Pr.~Lavrentyeva
13, 630090 Novosibirsk, Russia }

\date{November 6, 2009}

\begin{abstract}

Rates of depopulation by blackbody radiation (BBR) and effective
lifetimes of alkali-metal \textit{nS}, \textit{n}P, and
\textit{nD} Rydberg states have been calculated in a wide range of
principal quantum numbers $n \le 80$ at the ambient temperatures
of 77, 300 and 600~K. Quasiclassical formulas were used to
calculate the radial matrix elements of the dipole transitions
from Rydberg states. Good agreement of our numerical results with
the available theoretical and experimental data has been found. We
have also obtained simple analytical formulas for estimates of
effective lifetimes and BBR-induced depopulation rates, which well
agree with the numerical data.
\end{abstract}

\maketitle

\section{Introduction}

Accurate determination of effective lifetimes is important for
many theoretical and experimental studies of alkali-metal Rydberg
atoms. First measurements and calculations of radiative lifetimes
of Na \textit{nS} and \textit{nD} Rydberg states with $n \le 13$
were done by Gallagher et al.~\cite{Gallagher1975}. Later on it
has been shown that interaction of Rydberg atoms with blackbody
radiation (BBR) strongly affects the measured
lifetimes~\cite{Gallagher1979}. BBR-induced depopulation rates
were estimated by Cooke and Gallagher~\cite{Cooke1980} for sodium
Rydberg states with \textit{n}=18 and 19, and accurately
calculated by Farley and Wing~\cite{FarleyWing} for alkali-metal
\textit{nS}, \textit{nP}, and \textit{nD} Rydberg atoms with $n
\le 30 $ using Coulomb approximation~\cite{Zimmerman}. Radiative
lifetimes of sodium \textit{nS} and \textit{nD} Rydberg states
with 17$ \le $\textit{n$ \le $}28 were measured by Spencer et
al.~\cite{Kleppner1981} in a cooled environment in order to reduce
the influence of BBR. The temperature dependence of BBR-induced
depopulation rate was measured experimentally for the sodium
19\textit{S} state~\cite{Kleppner1982} and compared with numerical
calculations. Theodosiou~\cite{Theodosiou} performed the
model-potential calculations of the effective lifetimes of
alkali-metal Rydberg states with \textit{n}$ \le $21 for several
ambient temperatures in the range from 0 to 720~K. Later on,
model-potential calculations of radiative lifetimes for an
extended range of \textit{n} were done by He et al.~\cite{He}. For
most of alkali-metal Rydberg atoms, radiative lifetimes were
calculated up to \textit{n}=30, while for rubidium the
calculations were done up to \textit{n}=50. For determination of
effective lifetimes at 300~K and comparison with available
experimental results, the authors of Ref.~\cite{He} used the data
of Farley and Wing~\cite{FarleyWing}. Galvez et al.~\cite{Galvez1,
Galvez2} investigated the cascade of BBR-induced transitions from
the initially populated \textit{n}=24-29 states of Na, both
theoretically and experimentally.

Recent experimental studies of cold alkali-metal Rydberg atoms in
magneto-optical traps involved states with relatively high
principal quantum numbers $n>50$ \cite{Tanner}. Effective
lifetimes and BBR-induced depopulation rates for these states have
not been calculated yet, to the best of our knowledge. Commonly
used results of \cite{FarleyWing} and \cite{Theodosiou} were
limited by \textit{n}=21 and \textit{n}=30 respectively. We note
that experiments with cold Rydberg atoms are usually performed at
a room temperature of 300~K, when depopulation of Rydberg states
by blackbody radiation is the main source of the reduction in
radiative lifetimes as it was shown by Gallagher and
Cooke~\cite{Gallagher1979}. Recently, room-temperature
measurements of lifetimes of rubidium \textit{nS}, \textit{nP},
and \textit{nD} Rydberg atoms with \textit{n}=26-45 have been done
\cite{Marcassa65, Marcassa74} and discussed \cite{Tate,
Marcassa75}.

The present work is devoted to the calculations of the effective
lifetimes of \textit{nS}, \textit{nP}, and \textit{nD} Rydberg
states of alkali-metal Rydberg atoms with \textit{n$ \le $}80 at
the ambient temperatures of 77, 300 and 600~K. A simple
theoretical model describing spontaneous and BBR-induced
transitions between Rydberg states is discussed in Sec.~II.
Section~III is devoted to the calculations of the
temperature-dependent BBR-induced depopulation rates. The
analytical formulas for estimates of BBR-induced depopulation
rates \cite{Cooke1980} are modified to improve the agreement with
numerical results. In Section~IV the results of the numerical
calculations of effective lifetimes of Rydberg states are
presented and compared with available experimental and theoretical
data. Simple scaling laws for estimates of effective lifetimes are
obtained. Atomic units are used, unless specified otherwise.

\section{Spontaneous and BBR-induced transitions between Rydberg
states}

A simple model for calculation of effective lifetimes of Rydberg
states was developed by Gallagher and Cooke~\cite{Gallagher1979}.
The rate of a spontaneous transition between \textit{nL} and
$n'L'$ states is given by the Einstein coefficient:

\begin{equation}
\label{eq1}
A\left( {nL \to n'L'} \right) = \frac{{4\omega _{n{n}'}^{3}
}}{{3c^{3}}}\frac{{L_{max}} }{{2L + 1}}R^{2}\left( {nL \to n'L'} \right).
\end{equation}

\noindent Here $L_{max}$ is the largest of $L$ and $L'$, $R\left(
{nL \to n'L'} \right)$ is a radial matrix element of the electric
dipole transition, and $\omega _{nn'} = \left| {E_{nL} -
E_{{n}'{L}'}} \right|$ is a transition frequency, where $E_{nL} $
and $E_{{n}'{L}'} $ are energies of \textit{nL} and ${n}'{L}'$
states, respectively. Energies of the Rydberg states are expressed
through the effective quantum number $n_{eff} = n - \mu _{L} $,
where $\mu _{L} $ is a quantum defect of an \textit{nL} Rydberg
state: $E_{nL}=-1/(2n_{eff}^2)$.

The rate of BBR-induced transitions $W\left( {nL \to n'L'}
\right)$ is expressed through the effective number of BBR photons
per mode $\bar {n}_{\omega}  $, given by the Planck distribution
at temperature \textit{T},

\begin{equation}
\label{eq2} \bar {n}_{\omega}  = \frac{{1}}{{\mathrm{exp}\left(
{{{\omega _{n{n}'}} \mathord{\left/ {\vphantom {{\omega _{n{n}'}}
{kT}}} \right. \kern-\nulldelimiterspace} {kT}}} \right) - 1}},
\end{equation}

\noindent where \textit{k} is the Boltzmann constant, and through
the Einstein coefficient,

\begin{equation}
\label{eq3} W\left( {nL \to n'L'} \right) = A\left( {nL \to n'L'}
\right)\bar {n}_{\omega}.
\end{equation}

\noindent Radiative lifetime $\tau _{0} $ of a Rydberg state is
determined by the total rate of spontaneous transitions from
\textit{nL} state to all lower-lying states:

\begin{equation}
\label{eqq3} \frac{{1}}{{\tau _{0}} } = \Gamma _{0} =
\sum\limits_{E_{nL} > E_{{n}'{L}'} } {A\left( {nL \to {n}'{L}'}
\right)} .
\end{equation}

\noindent The total rate of BBR-induced depopulation can be
written in a similar form, taking into account transitions to both
lower and higher states:

\begin{equation}
\label{eq4} \Gamma_{BBR} = \sum\limits_{n'}A\left(nL \to n'L'
\right) \frac{1}{\mathrm{exp}\left( \omega_{nn'}/ kT \right) - 1}.
\end{equation}

\noindent Finally, effective lifetime of the \textit{nL} Rydberg
state is determined by the sum of the rates $\Gamma _{0} $ and
$\Gamma _{BBR}$ of spontaneous and BBR-induced $nL \to {n}'{L}'$
transitions, respectively:

\begin{equation}
\label{eq5} \frac{1}{\tau _{eff}} = \Gamma_{0} + \Gamma _{BBR} =
\frac{{1}}{{\tau _{0}} } + \frac{1}{\tau_{BBR}}.
\end{equation}

\noindent A calculation of effective lifetimes is thus reduced to
a calculation of the radial matrix elements $R\left( {nL \to n'L'}
\right)$. Exact analytical solution exists only for a hydrogen
atom \cite{BetheSalpeter}. For alkali-metal atoms, various
numerical methods were developed. The Hartree-Fock and
multiconfiguration-interaction methods require long calculation
time. The Coulomb approximation method was applied by Farley and
Wing~\cite{FarleyWing} and Spencer et al.~\cite{Kleppner1981} for
numerical calculations and provided good agreement with
experimental results. Theodosiou~\cite{Theodosiou} and He et
al.~\cite{He} used a method of model potential with different
atomic potential functions.

A quasiclassical approximation is most suitable for states with
large principal quantum numbers $n>20$. Therefore, in this paper
we used a quasiclassical method developed by Dyachkov and
Pankratov~\cite{Dyachkov} to calculate radial matrix elements.
This approach is helpful for calculations, where large number of
dipole transitions must be considered. The authors of
Ref.~\cite{Dyachkov} showed that by the optimal choice of the mean
energy of the atomic states, it is possible to extend
semiclassical approximation and make it applicable even for states
with low $n$. Quantum defects of \textit{nS}, \textit{nP}, and
\textit{nD} Rydberg states of Li, Na, K, Rb and Cs atoms were
taken from refs.~\cite{Li}-\cite{Cs2} and used as the input
parameters for the calculations. Quantum defects of \textit{nF}
states of Na, K and Rb Rydberg atoms, required to calculate the
effective lifetimes of \textit{nD} states, were taken from
ref.~\cite{K}. For Li and Cs the quantum defects of \textit{nF}
states were taken from~\cite{Li} and \cite{Cs1}.

For atoms in the ground and low-excited states with large
frequencies of transitions at \textit{T}=300~K one has
$n_{\omega}\ll 1$, and the rates of BBR-induced transitions are
small. Hence, for atoms in such states the interaction with
blackbody radiation can be neglected. The situation is different
for Rydberg states: at transition frequencies on the order of
10$^4$ cm$^{-1}$ one has $\bar {n}_{\omega} \sim 10$, and the rate
of BBR-induced transitions can be ten times larger than the rate
of the spontaneous decay to neighboring Rydberg states. Hence,
depopulation by BBR must be necessarily taken into account when
calculating the lifetimes of Rydberg states.

\section{BBR-induced depopulation of Rydberg states}

\begin{figure}
\label{Fig1}
\includegraphics[width=7cm]{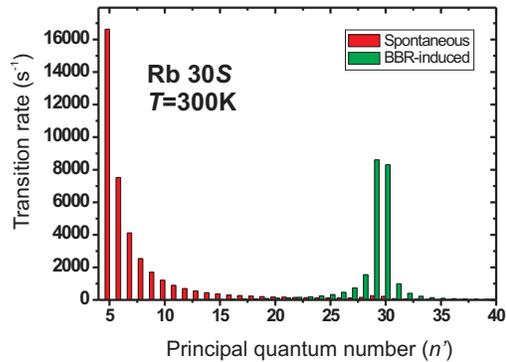}
\caption{(Color online) The rates of spontaneous and BBR-induced
transition from the rubidium 30\textit{S} state to ${n}'P$
states.}

\end{figure}


\begin{figure*}
\label{Fig2}
\includegraphics[width=15cm]{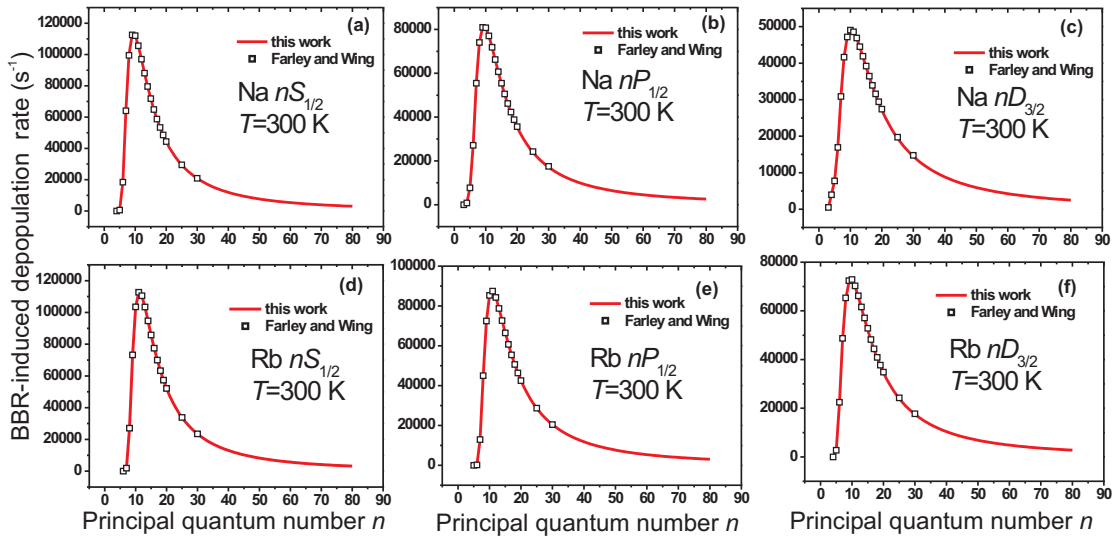}
\caption{(Color online) (a) Comparison of the numerically
calculated BBR-induced depopulation rates of (a) Na
\textit{nS}$_{1/2}$, (b) Na \textit{nP}$_{1/2}$, (c) Na
\textit{nD}$_{3/2}$, (d) Rb \textit{nS}$_{1/2}$, (e) Rb
\textit{nP}$_{1/2}$, (f) Rb \textit{nD}$_{3/2}$ Rydberg states
(this work) with numerical results of Farley and
Wing~\cite{FarleyWing}.}

\end{figure*}


Following the consideration of interaction of sodium Rydberg atoms
with BBR \cite{GallagherBook}, we  show numerically calculated
rates of spontaneous and BBR-induced transitions from the rubidium
30\textit{S} state to $n'P$ states in Fig.~1. For a given
\textit{n} spontaneous transitions occur predominantly to the
ground and low excited states, while blackbody radiation populates
mostly neighboring levels with $n' = n \pm 1$. Nevertheless, in
order to improve the precision of the numerical calculations of
BBR-induced depopulation rates we took into account transitions to
all lower states and to the upper states with $n' < n + 40$.
Omission of higher discrete states and continuum states reduces
the accuracy by less than 0.5\%~\cite{BBRIonization, Ovsiannikov}.

The range of validity of the commonly used theoretical model of
interaction of Rydberg atoms with blackbody radiation was
discussed by Farley and Wing~\cite{FarleyWing}. They  derived a
useful formula for determination of the critical values of
\textit{n} and \textit{T}, at which weak-field approximation
breaks down and interaction of the Rydberg electron with BBR
becomes as intensive, as Coulomb interaction of the electron with
the ionic core:

\begin{equation}
\label{eq6} kT \sim
\left(\frac{15}{32\pi^3}\right)^{1/4}\alpha^{5/4}\frac{m_e
c^2}{n^2}.
\end{equation}

\noindent Here $\alpha $ is the fine-structure constant, $m_{e} $
is the electron mass, and \textit{c} is the speed of light. For
\textit{T}=300~K, this formula gives \textit{n}=122, and for 600~K
the limit is \textit{n}=86. The condition of the breakdown of the
electric-dipole approximation is given by a simple formula,
$kTn^{2}\sim \alpha m_{e} c^{2}/3$, which yields \textit{n}=219
for 300~K and \textit{n}=155 for 600~K. The values of \textit{n}
studied in the present work are below the critical limits.

Figure~2 shows a comparison between our numerically calculated
BBR-induced depopulation rates of Na and Rb Rydberg states at
300~K with the results of Farley and Wing~\cite{FarleyWing}. It
demonstrates that both calculations agree very well.

\begin{figure*}
\label{Fig3}
\includegraphics[width=11cm]{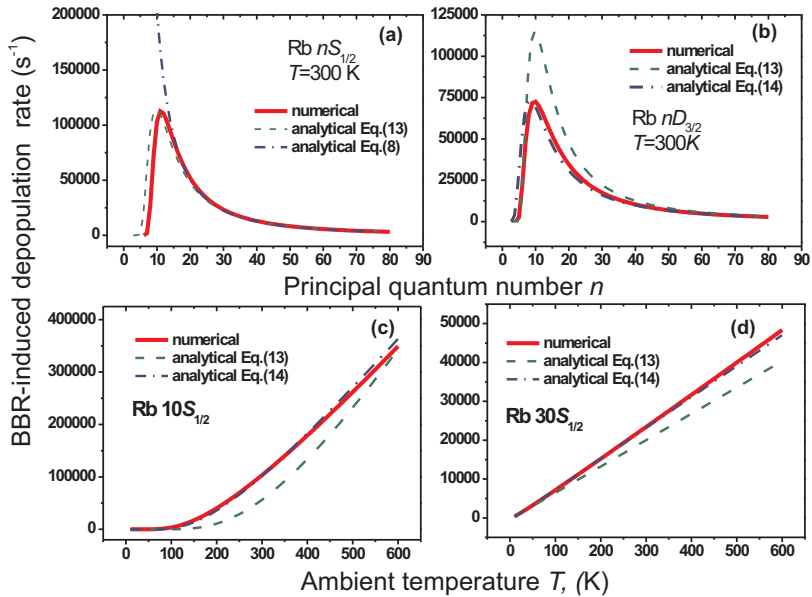}
\caption{(Color online) Comparison of our numerically calculated
by us depopulation rates of Rb \textit{nS} states at
\textit{T}=300~K with (a) Eq.~(\ref{eq7}) and Eq.~(\ref{eq11});
(b) Eq.~(\ref{eq11}) and Eq.~(\ref{eq12}), and comparison of our
numerically calculated temperature dependences of BBR-induced
depopulation rates of (c) Rb 10\textit{S} and (d) Rb 30\textit{S}
states with Eqs.~(\ref{eq11}) and (\ref{eq12}).}

\end{figure*}


Gallagher and Cooke~\cite{Cooke1980} used the sum rules to derive
a simple approximation for BBR-induced depopulation rate:

\begin{equation}
\label{eq7}
\Gamma _{BBR} = \frac{{4kT}}{{3c^{3}n_{eff}^{2}} }.
\end{equation}

\noindent Later on Farley and Wing~\cite{FarleyWing}  showed that
this formula overestimates the numerically calculated depopulation
rates, especially for low \textit{n}. Below we show how the
accuracy of Eq.~(\ref{eq7}) can be substantially improved.

Equation~(\ref{eq4}) can be rewritten as
\begin{equation}
\label{eq8} \Gamma _{BBR} = \frac{2}{c^3}\sum\limits_{n'} {\omega
_{nn'} \left| {f\left( {nL \to n'L'} \right)} \right|\left[
{\frac{{\omega _{nn'}} }{{\mathrm{exp}\left( {{\omega_{nn'}/
{kT}}} \right) - 1}}} \right]}.
\end{equation}

\noindent Here $f\left( {nL \to {n}'{L}'} \right)$ is the
oscillator strength:

\begin{equation}
\label{eq9} f\left( nL \to n'L' \right) = \frac{2}{3}\omega _{n'n}
R^2\left( nL \to n'L' \right).
\end{equation}

\noindent A principal contribution to the BBR depopulation rate is
caused by the transitions to neighboring levels with ${n}' = n \pm
1$ (see Fig.~1). One may note that an expression in the square
brackets in Eq.~(\ref{eq8}) is a slowly changing function of
$\omega _{nn'} $ for $n>15$ and it can be considered independently
of the other terms in the sum. For such states, $n'$ and
\textit{n} can be replaced by $n_{eff} $, and $\omega _{nn'} $ by
$n_{eff}^{-3} $. The remaining sum over the oscillator strengths
in Eq.~(\ref{eq8}) satisfies a sum rule \cite{BetheSalpeter}:

\begin{equation}
\label{eq9a} \sum\limits_{n'}\omega_{nn'}f\left(nL\to
n'L'\right)=\frac{2}{3 n_{eff}^2}.
\end{equation}

\noindent From Eqs.~(\ref{eq8})-(\ref{eq9a}) we obtain

\begin{equation}
\label{eq10} \Gamma _{BBR} = \frac{4}{3n_{eff}^{5}
c^{3}}\;\frac{1}{\mathrm{exp}\left[ 1/(n_{eff}^{3} kT ) \right] -
1}.
\end{equation}

\noindent For large \textit{n}, Eq.~(\ref{eq10}) can be expanded
and coincides with Eq.~(\ref{eq7}). It is convenient to rewrite it
in the units of s$^{-1}$, taking the temperature in kelvin:

\begin{equation}
\label{eq11} \Gamma _{BBR} = \frac{1}{n_{eff}^5} \;\frac{2.14
\times 10^{10}} {\mathrm{exp}\left[ 315780 /(n_{eff}^3 T ) \right]
- 1}\left[ {s^{ - 1}} \right].
\end{equation}

\noindent A comparison of Eqs.~(\ref{eq11}) and (\ref{eq7}) with
the numerical results is shown in Fig.~3(a). It is seen that
Eq.~(\ref{eq11}) correctly describes the shape of the numerically
calculated dependence, and for higher \textit{n} it yields the
results identical to Eq.~(\ref{eq7}). However, for \textit{nP} and
\textit{nD} states [Fig.~3(b)] Eq.~(\ref{eq11}) overestimates the
numerically calculated values at intermediate $n\sim 20$. We have
found that a better approximation of the numerical results for
large \textit{n} can be obtained by introducing the fitting
parameters \textit{A}, \textit{B}, \textit{C}, and \textit{D} in
Eq.~(\ref{eq11}):

\begin{equation}
\label{eq12} \Gamma _{BBR} = \frac{A}{n_{eff}^D}\;\frac{2.14
\times 10^{10}}{\mathrm{exp}\left[ 315780 \times B /( n_{eff}^C
T)\right] - 1}\left[ {s^{ - 1}} \right].
\end{equation}

\begin{figure*}
\label{Fig4}
\includegraphics[width=15cm]{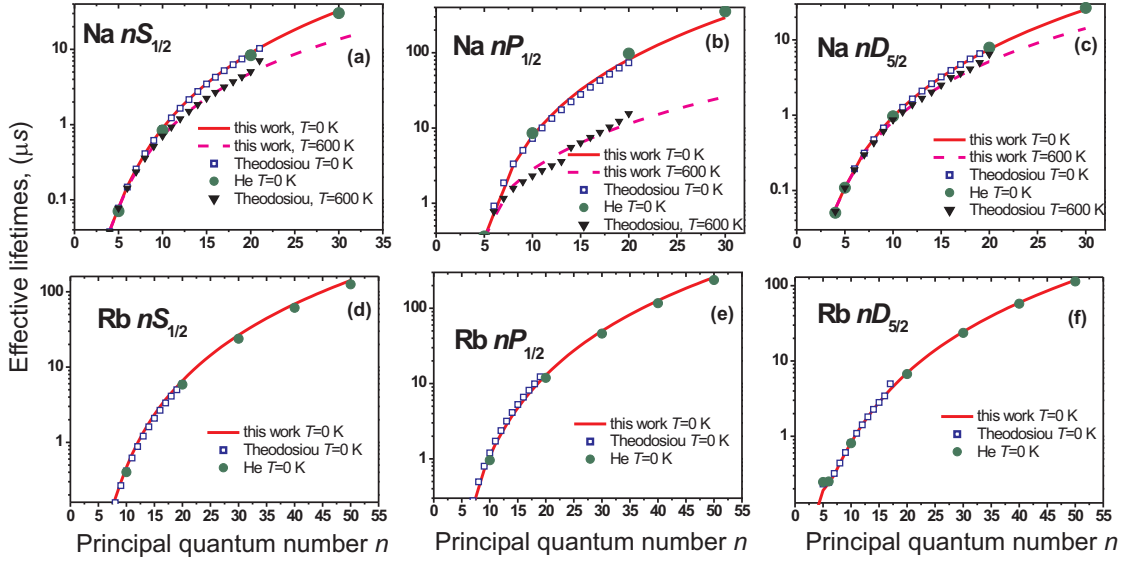}
\caption{(Color online) Comparison of the numerically calculated
radiative and effective lifetimes of (a) Na \textit{nS} , (b)
\textit{nP}, (c) \textit{nD}, and (d) Rb \textit{nS}, (e)
\textit{nP}, and (f) \textit{nD} Rydberg states with available
numerical data \cite{He, Theodosiou}.}

\end{figure*}


\begin{table*}
\caption{Scaling coefficients \textit{A, B, C, D} in
Eq.(\ref{eq12})}
\begin{tabular*}{\textwidth}{@{\extracolsep{\fill}} p{15pt}| p{30pt} p{30pt}
p{30pt}p{30pt}| p{30pt} p{30pt} p{30pt} p{30pt}| p{30pt} p{30pt}
p{30pt} p{30pt} }
 \hline \hline &  \multicolumn{4}{c|}{{\textbf{\textit{S}}$_{1/2}$}} &
\multicolumn{4}{c|}{\textbf{\textit{P}}$_{1/2}$} &
\multicolumn{4}{c}{\textbf{\textit{D}}$_{3/2}$}  \\ \cline{6-13}&
 \multicolumn{4}{c|}{}&
 \multicolumn{4}{c|}{\textbf{\textit{P}}$_{3/2}$} &
\multicolumn{4}{c}{\textbf{\textit{D}}$_{5/2}$}

\\ \hline &
\textbf{A}& \textbf{B}& \textbf{C}& \textbf{D}& \textbf{A}&
\textbf{B}& \textbf{C}& \textbf{D}& \textbf{A}& \textbf{B}&
\textbf{C}& \textbf{D} \\

\hline {Li}& 0.051& 0.097& 1.991&3.852& 0.040& 0.078& 1.712&
3.610& 0.058& 0.148& 1.934& 3.783\\

\hline {Na}&0.138& 0.259& 2.587& 4.446& 0.074& 0.117& 2.032&
3.977& 0.058& 0.109& 1.816& 3.724 \\ \cline{6-13}
 & & & & & 0.074&0.117& 2.033& 3.978& 0.058& 0.109& 1.816& 3.724\\

\hline {K}&{0.123}& {0.232}&{2.522}& {4.379}& 0.118& 0.257& 2.600&
4.421& 0.044& 0.104& 2.003& 3.831 \\ \cline{6-13}
 & & & & &0.105&0.236&2.568&4.382&0.044&0.103&2.002&3.830 \\

\hline {Rb}&{0.134}& {0.251}& {2.567}& {4.426}& 0.053&0.128&
2.183& 3.989&0.033& 0.084& 1.912& 3.716
\\\cline{6-13} & & & & &0.046&0.109&2.085&3.901&0.032&0.082&1.898&3.703 \\

\hline {Cs}& {0.123}&{0.231}& {2.517}&{4.375}& 0.041& 0.072&
1.693& 3.607& 0.038& 0.076& 1.790& 3.656 \\ \cline{6-13} & & & & &
0.038& 0.056& 1.552& 3.505& 0.036& 0.073& 1.770& 3.636 \\ \hline
\hline
\end{tabular*}
\end{table*}

\noindent The values of \textit{A}, \textit{B}, \textit{C}, and
\textit{D}, obtained from the best fit of the numerical results at
\textit{T}=300~K to Eq.~(\ref{eq12}), are summarized in Table~I.
For Li Rydberg states the fine structure is not considered, as it
does not affect the results. Figure~3 shows that Eq.~(\ref{eq12})
provides  better agreement with the numerical calculations than
Eq.~(\ref{eq11}), both for dependences on the principal quantum
number [Fig.~3(b)] and on the ambient temperature
[Figs.~3(c),(d)].

\begin{figure*}
\label{Fig5}
\includegraphics[width=15cm]{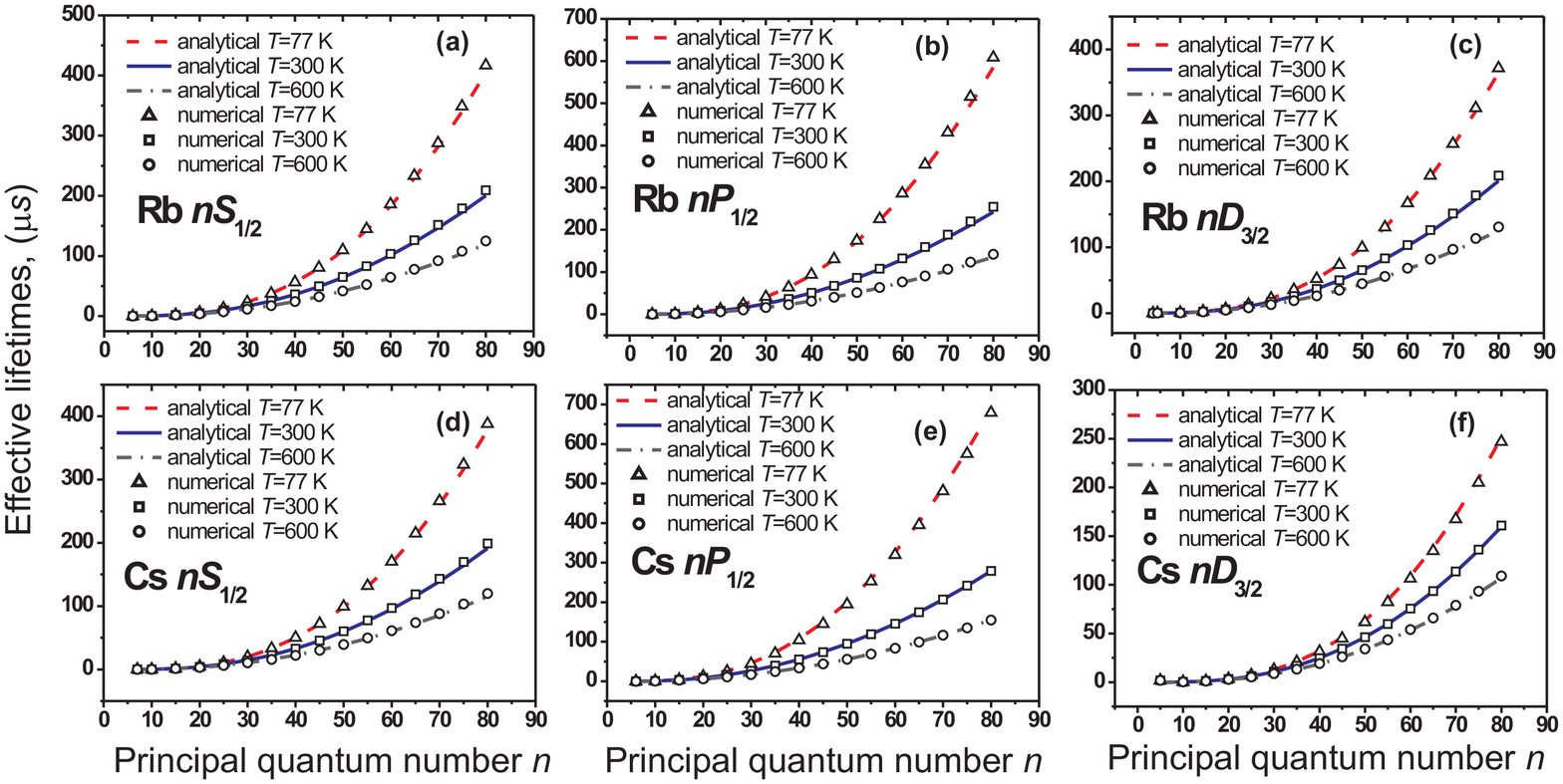}
\caption{(Color online)  Comparison of the numerically calculated
by us effective lifetimes of Rb \textit{nS} (a), \textit{nP} (b),
\textit{nD} (c) and Cs \textit{nS}(d), \textit{nP} (e) and
\textit{nD} (f) Rydberg states at the ambient temperatures
\textit{T}=77, 300 and 600~K with Eq.~(\ref{eq14}).}

\end{figure*}

Basically, all scaling laws are better for interpolation than for
extrapolation of the precise numerical data. It is difficult to
obtain an accuracy better than 50\% for $n\sim 80$, using the only
available results of Farley and Wing~\cite{FarleyWing} as the
input data for scaling with Eq.~(\ref{eq7}).  From Fig.~3(a) one
may see that Eq.~(\ref{eq11}) gives the completely different and
more correct results for $n<15$, compared to the commonly used
Eq.~(\ref{eq7}). Furthermore, we have shown that for more accurate
analytical calculations of the depopulation rates, scaling
coefficients should be introduced, as we did in Eq.~(\ref{eq12})
for the first time. Although the data of Ref.~\cite{FarleyWing}
with Eq.~(\ref{eq12}) can be used to obtain the scaling with 10\%
accuracy at $n=50$ and 18\% accuracy at $n=80$, compared to our
numerical calculations, the best fit of our numerical results in
an extended range of $10<n<80$ allows us to obtain more accurate
values of the coefficients in Eq.~(\ref{eq12}), providing
accuracies better than 3.5\% for $n\sim50$ and better than 8\% for
$n\sim 80$. This improvement can be important for comparison with
precise experimental data.

\section{Effective lifetimes of Rydberg states}

We have numerically calculated the radiative and effective
lifetimes of \textit{nS}, \textit{nP}, and \textit{nD}
alkali-metal Rydberg states using Eqs.~(\ref{eq1})-(\ref{eq5}). A
semi-empirical formula is commonly used for approximation of
numerical results on radiative lifetimes $\tau _{0}$
\cite{Gounand}:

\begin{equation}
\label{eq13}
\tau _{0} = \tau _{s} n_{eff}^{\delta}  \left[ {ns} \right].
\end{equation}

\noindent The coefficients $\tau_s$ and $\delta$ have been
obtained from the best fit of our numerical results and are
summarized in Table~II. The radiative lifetimes $\tau_0$ of
alkali-metal Rydberg states calculated by us are compared with the
available theoretical data \cite{Theodosiou, He} in Table~III and
in Fig.~4 (for Na and Rb). Good agreement between the three data
sets is observed.

\begin{table*}
\caption{ Scaling coefficients $\tau_s$ and $\delta$ in
Eq.(\ref{eq13})}
\begin{tabular*}{\textwidth}{@{\extracolsep{\fill}}p{31pt}|p{43pt}p{43pt}|p{43pt}p{43pt}|p{43pt}
p{43pt}|p{43pt}p{43pt}|p{43pt}p{43pt}} \hline \hline &
\multicolumn{2}{p{87pt}|}{\textit{S}$_{1/2}$} &
\multicolumn{2}{p{87pt}|}{\textit{P}$_{1/2}$} &
\multicolumn{2}{p{87pt}|}{\textit{P}$_{3/2}$} &
\multicolumn{2}{p{87pt}|}{\textit{D}$_{3/2}$} &
\multicolumn{2}{p{87pt}}{\textit{D}$_{5/2}$}  \\ \hline {}&
$\tau_s$ &  $\delta$ & $\tau_s $ \textbf{}&
 $\delta $ \textbf{}&
 $\tau _{s} $ \textbf{}&
 $\delta $ \textbf{}&
 $\tau _{s} $ \textbf{}&
 $\delta $ \textbf{}&
 $\tau _{s} $ \textbf{}&
 $\delta $ \textbf{} \\
\hline Li& 0.8431& 2.9936& 2.8807& 2.9861& & & 0.4781& 2.9963& &
 \\
\hline Na& 1.3698& 3.0018& 12.052& 2.9969& 11.862& 2.9972& 0.9555&
2.9971& 0.9560& 2.9971 \\ \hline K& 3.6163& 2.9966& 3.6415&
3.0009& 3.6163& 2.9966& 2.3168& 2.9829& 2.2972& 2.9831 \\ \hline
Rb& 1.368& 3.0008& 2.4360& 2.9989& 2.2214& 3.0026& 1.0761& 2.9898&
1.0687& 2.9897 \\ \hline Cs& 1.2926& 3.0005& 2.9921& 2.9892&
3.2849& 2.9875& 0.6580& 2.9944& 0.6681& 2.9941 \\ \hline \hline
\end{tabular*}
\end{table*}


\begin{figure*}
\label{Fig6}
\includegraphics[width=15cm]{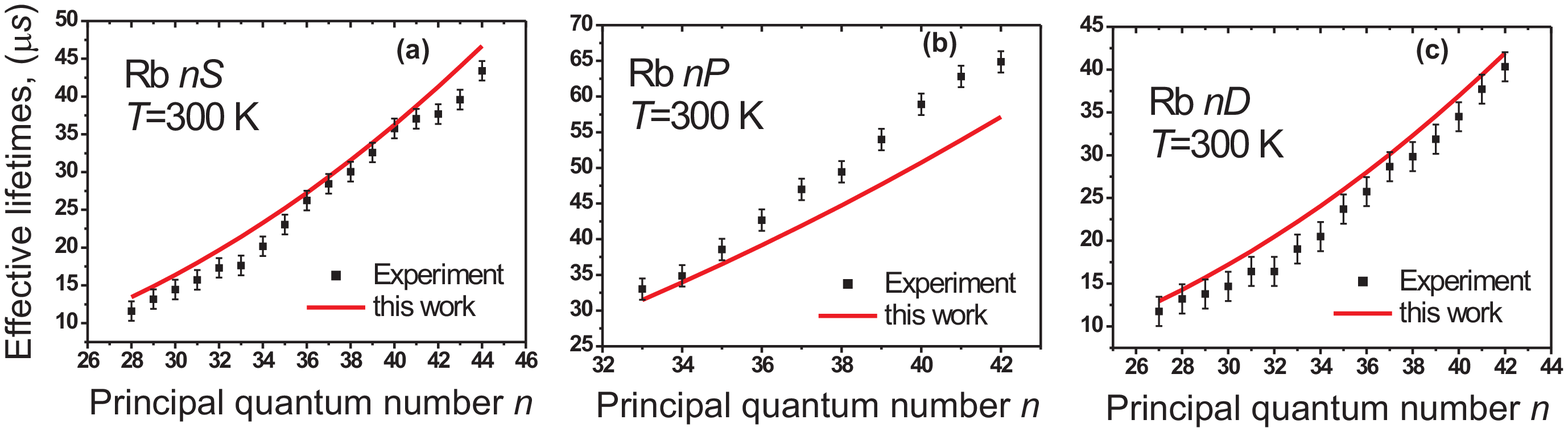}
\caption{ Comparison of the calculated effective lifetimes of Rb
\textit{nS} (a), \textit{nP} (b) and \textit{nD} (c) Rydberg
states with the experiments \cite{Marcassa65, Marcassa74} at
\textit{T}=300~K.}

\end{figure*}


A combination of Eqs.~(\ref{eq12}) and (\ref{eq13}) with
Eq.~(\ref{eq5}) can be used for estimates of the effective
lifetimes of alkali-metal Rydberg states at a given temperature:

\begin{eqnarray}
\label{eq14} \tau_{eff} =\left(\frac{{1}}{{\tau_s
n_{eff}^{\delta}}}\,+ \right. \qquad \qquad \qquad \qquad  \qquad
\qquad \qquad \qquad &&
\\+ \left. \frac{A}{n_{eff}^D}\;\frac{21.4}{\mathrm{exp}\left[
315780 \times B / ( n_{eff}^C T ) \right] - 1}\right
)^{-1}\left[ns\right]. &&\nonumber
\end{eqnarray}

\begin{table*}
\caption{Comparison of the calculated radiative lifetimes $\tau
_{0} $ (in nanoseconds) of alkali-metal Rydberg states with
available theoretical data.}
\begin{tabular*}{\textwidth}{@{\extracolsep{\fill}}p{15pt}p{23pt}|p{45pt}p{45pt}p{45pt}|p{45pt}p{45pt}p{45pt}|p{45pt}
p{45pt}p{45pt}} \hline \hline &{\textit{n}}&
\multicolumn{3}{p{135pt}|}{\textit{S}$_{1/2}$} &
\multicolumn{3}{p{135pt}|}{\textit{P}$_{1/2}$} &
\multicolumn{3}{p{153pt}}{\textit{D}$_{5/2}$}  \\

\hline & & This work& \cite{Theodosiou}& \cite{He}& This work&
\cite{Theodosiou}& \cite{He}& This work& \cite{Theodosiou}&
\cite{He}
\\ \hline {Li}& 3& 28.51& 30.04& --& 181.3& 211.9& --& 13.71&
14.64& -- \\ \cline{2-11} &4& 52.68& 56.29& --& 411.4& 391.2& --&
24.32&33.49&-- \\ \cline{2-11}
&5&99.05&102.5&--&704.1&610.3&--&62.39&63.89&-- \\\cline{2-11} &
10&757.1& 783.4& --& 3128& 3328& --& 457.4& 487.3& --
\\ \cline{2-11} &15&2606&2695&--&9787&10280&--&1621&1621&-- \\
\cline{2-11}
 &20&6263&6457&--&22610&23520&--&3811&
3818&-- \\ \hline{Na}& 3& --& --& --&17.005& 16.140& 14.900&
19.887& 19.470& 17.300 \\ \cline{2-11}
 &4&39.074&37.710&32.900&115.07&107.19&101.00&56.719&52.500&50.600 \\
\cline{2-11}
 &5&78.120&77.640&70.700&344.32&369.86&362.00&109.74&108.87&109.00 \\\cline{2-11}
 &10&895.24&888.23&838.00&7849.5&7264.7&8560.0&954.77&956.97&968.00 \\\cline{2-11}
 &15&3506.0&3456.3&--&32176&27845&--&3198.3&3218.2&-- \\
\cline{2-11}
 &20&8920.8&8804.2&8350.0&81359&73653&97500&7539.5&7691.0&7910.0 \\\cline{2-11}
 &30&32397&--&30300&291860& &
287530&25391& --& 26700 \\ \hline {K}& 4& --& --& --& 27.90&
27.51& 22.40& 281.4& 291.2& 269.0 \\ \cline{2-11}
 &5&46.30&
46.50&
35.80&
134.3&
127.1&
117.0&
627.2&
769.6&
502.0 \\
\cline{2-11}
 &
6&
86.40&
87.12&
--&
298.1&
321.7&
--&
814.8&
1169&
-- \\
\cline{2-11}
 &
10&
626.0&
623.3&
553.0&
2202&
2346&
2330&
2601&
3492&
1980 \\
\cline{2-11}
 &
15&
2704&
2644&
--&
8798&
9415&
--&
7589&
10260&
-- \\
\cline{2-11}
 &
20&
7222&
6846&
6430&
22730&
23070&
24400&
17410&
--&
13600 \\
\cline{2-11}
 &
30& 27000& --& 24500& 82700& & 89900& 57490& --& 44800 \\ \hline
{Rb}& 5& --& --& --& 29.233& 27.040& 20.600& 190.88& 231.78&
247.00 \\ \cline{2-11}
 &
6&
51.554&
45.210&
38.200&
115.07&
124.03&
83.900&
237.21&
243.72&
251.00 \\
\cline{2-11}
 &
10&
463.92&
417.84&
402.00&
1076.8&
1201.4&
960.00&
820.06&
822.10&
807.00 \\
\cline{2-11}
 &
15&
2339.8&
2092.3&
--&
4742.4&
5269.0&
--&
2828.5&
2809.4&
-- \\
\cline{2-11}
 &
20&
6576.8&
5991.4&
5870.0&
12903&
--&
11900&
6939.3&
--&
6700.0 \\
\cline{2-11}
 &
30&
26594&
--&
23800&
50010&
--&
46100&
24471&
--&
23600 \\
\cline{2-11}
 &
40&
68751&
--&
61400&
126540&
--&
117000&
59516&
--&
57700 \\
\cline{2-11}
 &
50& 141310& --& 126000& 256280& --& 238000& 118210& --& 114000 \\
\hline {Cs}& 5& --& --& --& --& --& --& 1402& 1283& 713.0 \\
\cline{2-11}
 &
6&
--&
--&
--&
32.59&
33.66&
24.70&
61.58&
58.39&
63.60 \\
\cline{2-11}
 &
7&
55.15&
48.17&
40.00&
130.9&
159.3&
95.20&
93.07&
88.06&
91.30 \\
\cline{2-11}
 &
10&
293.2&
274.0&
257.0&
893.6&
1051&
741.0&
317.6&
315.3&
322.0 \\
\cline{2-11}
 &
15&
1746&
1618&
--&
4650&
5318&
--&
1357&
1336&
-- \\
\cline{2-11}
 &
20&
5263&
--&
4840&
13290&
--&
11700&
3617&
--&
3650 \\
\cline{2-11}
 &
30& 22640& --& 20900& 53370& --& 47800& 13750& --& 14000 \\ \hline
\hline
\end{tabular*}
\end{table*}

\noindent In Figs.~4(a)-4(c) the effective lifetimes $\tau _{eff}
$ calculated for Na at \textit{T}=600~K are compared with the
results of Theodosiou \cite{Theodosiou}. Satisfactory agreement
with our numerical calculations is observed for low $n<20$. At
$n\sim 20$ the results of Theodosiou start to deviate from both a
smooth dependence and our calculations. The deviation from the
smooth dependence was probably caused by the difficulties of
numerical integration of rapidly oscillating wave functions of
Rydberg states with large \textit{n}.

A comparison of numerically calculated effective lifetimes of Rb
and Cs \textit{nS}, \textit{nP}, and \textit{nD} Rydberg states at
the ambient temperatures of 77, 300, and 600~K with
Eq.~(\ref{eq14}) is shown in Fig.~5. Good agreement is observed
for all states and \textit{n}. We have found that in the range
$15<n<80$ Eq.~(\ref{eq14}) can be used for estimates of the
effective lifetimes with the accuracy better than 5\%. This
indicates that Eq.~(\ref{eq14}) has a good precision and can be
used for prompt analytical estimates of the effective lifetimes of
all alkali-metal Rydberg states.

We have also compared the calculated effective lifetimes of Rb
Rydberg states with those from the experiments of de~Oliveira and
co-workers~\cite{Marcassa65, Marcassa74}. The data for \textit{nS}
and \textit{nD} states were taken from the recent work
\cite{Marcassa74}, while for \textit{nP} states we used the
results of measurements published in the earlier work
\cite{Marcassa65}. In the latter paper the authors used an
uncommon definition of effective lifetimes, and their results must
be divided by a factor 2, as mentioned in Ref.~\cite{Marcassa74}.
Satisfactory agreement between the experiment and our calculations
is observed (see Fig.~6). The theoretical curves for \textit{nS}
and \textit{nD} states go slightly higher than the experimental
points, while for the \textit{nP} states they go below
experimental points. A reason for this discrepancy is unclear. On
the one hand, it can be caused by inaccuracy of the quasiclassical
model applied for calculations of radial matrix elements of
transitions to lower states with \textit{n}$\sim $3-5, which make
the principal contribution to radiative lifetimes (see Fig.~1). On
the other hand, we have checked that our quasiclassical model
gives  better agreement with the experiment and model-potential
calculations \cite{Theodosiou, He}, than the commonly used Coulomb
approximation method \cite{Zimmerman}. For \textit{nP} states, the
disagreement between theory and the experiment is more
significant, and we may conclude that more accurate experimental
measurements would be of great interest.

Finally, we present the results of our numerical calculations of
the radiative and effective lifetimes of \textit{nS}, \textit{nP},
and \textit{nD} alkali-metal Rydberg states with \textit{n}=10-80
at the ambient temperatures of \textit{T}=77, 300, and 600~K in
Tables~IV-VIII.

\section{ Conclusion.}

We reported the results of our numerical calculations of
BBR-induced depopulation rates and effective lifetimes of
\textit{nS}, \textit{nP}, and \textit{nD} alkali-metal Rydberg
states, which were extended to higher principal quantum numbers
$n\leq 80$ in comparison with previous publications
\cite{FarleyWing, He, Theodosiou}. Good agreement of the
calculated BBR depopulation rates with the results of Farley and
Wing \cite{FarleyWing} for $n \le 30$ proves the validity of the
quasiclassical methods \cite{Dyachkov} used in the present work.
Our results of numerical calculations of spontaneous radiative
lifetimes are also consistent with the previous theoretical works
\cite{He, Theodosiou}. We have also obtained satisfactory
agreement with the experimental measurements \cite{Marcassa65,
Marcassa74} of the effective lifetimes of Rb \textit{nS},
\textit{nP}, and \textit{nD} states with \textit{n}=26-45.
Nevertheless, the remaining discrepancies between experiment and
theory indicate that new experimental measurements for
alkali-metal Rydberg states in a wider range of \textit{n} would
be of great interest.

We have also derived an improved analytical formula
[Eq.~(\ref{eq11})] for prompt estimates of the BBR-induced
depopulation rates, which better agrees with the results of
numerical calculations for lower \textit{n}, than the commonly
used Eq.~(\ref{eq7}) \cite{Cooke1980}. The simple scaling laws
[Eqs.~(\ref{eq12}) and (\ref{eq14})] based on Eq.~(\ref{eq11}) can
be used for accurate approximation of the results of numerical
calculations of BBR-induced depopulation rates and effective
lifetimes.

\begin{acknowledgments}

This work was supported by the Russian Academy of Sciences, RFBR,
Bilateral RFBR-EINSTEIN project, and Dynasty Foundation.

\end{acknowledgments}

\newpage

\begin{table*}
\caption{Effective lifetimes $\tau _{eff} \left( {\mu s} \right)$
of Li \textit{nS}, \textit{nP}, and \textit{nD} Rydberg states.}

\begin{tabular*}{\textwidth}{@{\extracolsep{\fill}}p{10pt}|p{35pt}p{35pt}p{35pt}|p{35pt}p{35pt}p{35pt}|
p{35pt}p{35pt}p{35pt}| p{35pt}p{35pt}p{35pt}} \hline \hline &
\multicolumn{3}{p{117pt}|}{\textit{T}=0~K} &
\multicolumn{3}{p{114pt}|}{\textit{T}=77~K} &
\multicolumn{3}{p{114pt}|}{\textit{T}=300~K} &
\multicolumn{3}{p{114pt}}{\textit{T}=600~K}  \\ \hline \textit{n}&
\textit{S}& \textit{P}& \textit{D}& \textit{S}& \textit{P}&
\textit{D}& \textit{S}& \textit{P}& \textit{D}& \textit{S}&
\textit{P}& \textit{D} \\ \hline 10& 0.757& 3.128& 0.457& 0.753&
3.094& 0.457& 0.711& 2.679& 0.449& 0.650& 2.124& 0.432 \\ \hline
15& 2.606& 9.787& 1.621& 2.544& 9.290& 1.612& 2.274& 7.033& 1.536&
1.962& 5.119& 1.425 \\ \hline 20& 6.263& 22.61& 3.811& 5.991&
20.37& 3.757& 5.097& 13.94& 3.484& 4.193& 9.599& 3.142 \\ \hline
25& 12.32& 43.58& 7.412& 11.55& 37.21& 7.236& 9.396& 23.55& 6.545&
7.427& 15.55& 5.754 \\ \hline 30& 21.43& 74.45& 12.77& 19.69&
60.41& 12.35& 15.36& 35.86& 10.91& 11.74& 22.95& 9.364 \\ \hline
35& 34.20& 117.5& 20.25& 30.82& 90.80& 19.38& 23.13& 51.01& 16.74&
17.16& 31.82& 14.06 \\ \hline 40& 51.24& 174.8& 30.19& 45.30&
129.0& 28.61& 32.80& 69.03& 24.18& 23.72& 42.16& 19.90 \\ \hline
45& 73.16& 248.3& 42.96& 63.47& 175.4& 40.32& 44.46& 89.89& 33.37&
31.44& 53.95& 26.94 \\ \hline 50& 100.6& 339.9& 58.89& 85.68&
230.2& 54.74& 58.18& 113.6& 44.40& 40.33& 67.17& 35.21 \\ \hline
55& 133.6& 451.8& 78.35& 111.8& 294.0& 72.14& 73.83& 140.2& 57.38&
50.32& 81.83& 44.76 \\ \hline 60& 173.7& 586.1& 101.7& 142.8&
366.8& 92.75& 91.78& 169.6& 72.39& 61.57& 97.93& 55.60 \\ \hline
65& 221.2& 744.4& 129.2& 178.7& 449.0& 116.8& 111.9& 201.9& 89.50&
74.03& 115.5& 67.77 \\ \hline 70& 276.6& 929.3& 161.4& 219.7&
540.7& 144.5& 134.3& 237.0& 108.8& 87.68& 134.4& 81.28 \\ \hline
75& 340.5& 1142& 198.5& 266.0& 642.1& 176.1& 158.9& 275.0& 130.3&
102.5& 154.8& 96.14 \\ \hline 80& 413.7& 1386& 240.9& 317.8&
753.4& 211.8& 185.8& 315.8& 154.2& 118.6& 176.6& 112.4 \\ \hline
\hline
\end{tabular*}
\end{table*}

\bigskip

\begin{table*}
\caption{Effective lifetimes $\tau _{eff} \left( {\mu s} \right)$
of Na \textit{nS}, \textit{nP}, and \textit{nD} Rydberg states.}
\begin{tabular*}{\textwidth}{@{\extracolsep{\fill}}p{10pt}|p{35pt}p{35pt}p{35pt}|p{35pt}p{35pt}p{35pt}|p{35pt}p{35pt}p{35pt}|
p{35pt}p{35pt}p{35pt}} \hline \hline
 \textit{}& \multicolumn{3}{p{117pt}|}{\textit{T}=0~K} &
\multicolumn{3}{p{114pt}|}{\textit{T}=77~K} &
\multicolumn{3}{p{114pt}|}{\textit{T}=300~K} &
\multicolumn{3}{p{114pt}}{\textit{T}=600~K}  \\ \hline
{\textit{n}}& \textit{S}$_{1/2}$& \textit{P}$_{1/2}$&
\textit{D}$_{3/2}$& \textit{S}$_{1/2}$& \textit{P}$_{1/2}$&
\textit{D}$_{3/2}$& \textit{S}$_{1/2}$& \textit{P}$_{1/2}$&
\textit{D}$_{3/2}$& \textit{S}$_{1/2}$& \textit{P}$_{1/2}$&
\textit{D}$_{3/2}$ \\ \cline{2-13}
 &
& \textit{P}$_{3/2}$& \textit{D}$_{5/2}$& & \textit{P}$_{3/2}$&
\textit{D}$_{5/2}$& & \textit{P}$_{3/2}$& \textit{D}$_{5/2}$& &
\textit{P}$_{3/2}$& \textit{D}$_{5/2}$ \\ \hline 10& 0.8952&
7.8495& 0.9542& 0.8905& 7.5702& 0.9524& 0.8135& 4.8014& 0.9115&
0.7068& 2.8247& 0.8333 \\ \hline --& --& 7.7377& 0.9548& --&
7.4653& 0.9530& --& 4.7562& 0.9120& --& 2.8078& 0.8337 \\ \hline
15& 3.5060& 32.176& 3.1970& 3.3671& 25.594& 3.1517& 2.8010&
11.585& 2.8433& 2.2491& 6.3604& 2.4529 \\ \hline --& --& 31.696&
3.1983& --& 25.279& 3.1529& --& 11.514& 2.8442& --& 6.3366& 2.4535
\\ \hline 20& 8.9208& 81.359& 7.5357& 8.2551& 52.902& 7.2957&
6.3928& 20.989& 6.2556& 4.8533& 11.364& 5.1582 \\ \hline --& --&
80.146& 7.5395& --& 52.363& 7.2992& --& 20.894& 6.2580& --&
11.332& 5.1597 \\ \hline 25& 18.209& 164.29& 14.717& 16.272&
89.895& 13.967& 11.853& 33.333& 11.444& 8.6116& 17.880& 9.0771 \\
\hline --& --& 156.2& 14.725& --& 89.119& 13.975& --& 33.212&
11.448& --& 17.839& 9.0797 \\ \hline 30& 32.397& 291.86& 25.376&
28.016& 137.57& 23.609& 19.343& 48.712& 18.565& 13.564& 25.912&
14.246 \\ \hline --& --& 287.53& 25.391& --& 136.54& 23.621& --&
48.563& 18.572& --& 25.862& 14.249 \\ \hline 35& 52.520& 474.31&
40.241& 44.024& 196.30& 36.708& 28.980& 67.105& 27.789& 19.736&
35.444& 20.716 \\ \hline --& --& 467.29& 40.265& --& 195.00&
36.727& --& 66.928& 27.799& --& 35.385& 20.721 \\ \hline 40&
79.618& 714.32& 60.488& 64.778& 265.27& 54.075& 40.851& 88.375&
39.435& 27.147& 46.433& 28.620 \\ \hline --& --& 703.77& 60.525&
--& 263.69& 54.104& --& 88.167& 39.449& --& 46.364& 28.626 \\
\hline 45& 114.71& 1025.1& 86.073& 90.697& 345.46& 75.491& 55.017&
112.61& 53.250& 35.804& 58.901& 37.787 \\ \hline --& --& 1010.0&
86.125& --& 343.59& 75.530& --& 112.38& 53.268& --& 58.822& 37.795
\\ \hline 50& 158.86& 1413.3& 118.03& 122.18& 436.70& 101.59&
71.530& 139.78& 69.458& 45.720& 72.836& 48.310 \\ \hline --& --&
1392.5& 118.10& --& 434.53& 101.65& --& 139.51& 69.480& --&
72.746& 48.320 \\ \hline 55& 213.09& 1890.3& 157.02& 159.56&
539.32& 132.70& 90.430& 169.90& 88.105& 56.900& 88.241& 60.193 \\
\hline --& --& 1862.5& 157.12& --& 536.83& 132.77& --& 169.60&
88.132& --& 88.140& 60.204 \\ \hline 60& 278.44& 2462.6& 203.80&
203.14& 653.15& 169.14& 111.74& 202.95& 109.25& 69.346& 105.11&
73.447 \\ \hline --& --& 2426.5& 203.93& --& 650.33& 169.23& --&
202.62& 109.28& --& 105.00& 73.459 \\ \hline 65& 355.95& 3141.3&
259.05& 253.22& 778.36& 211.21& 135.49& 238.94& 132.92& 83.063&
123.44& 88.070 \\ \hline --& --& 3095.3& 259.21& --& 775.21&
211.31& --& 238.57& 132.96& --& 123.32& 88.084 \\ \hline 70&
446.64& 3933.2& 323.48& 310.03& 914.82& 259.18& 161.70& 277.84&
159.16& 98.053& 143.24& 104.07 \\ \hline --& --& 3875.6& 323.68&
--& 911.34& 259.31& --& 277.44& 159.20& --& 143.10& 104.09 \\
\hline 75& 551.57& 4846.4& 397.79& 373.82& 1062.5& 313.30& 190.38&
319.66& 187.98& 114.32& 164.48& 121.45 \\ \hline --& --& 4775.5&
398.04& --& 1058.7& 313.45& --& 319.22& 188.03& --& 164.34& 121.46
\\ \hline 80& 671.75& 5893.0& 482.71& 444.78& 1221.7& 373.82&
221.55& 364.40& 219.42& 131.86& 187.20& 140.20 \\ \hline & &
5806.9& 483.01& & 1217.5& 373.99& & 363.93& 219.47& --& 187.04&
140.22 \\ \hline \hline
\end{tabular*}
\end{table*}

\bigskip

\begin{table*}
\caption{Effective lifetimes $\tau _{eff} \left( {\mu s} \right)$
of K \textit{nS}, \textit{nP}, and \textit{nD} Rydberg states.}
\begin{tabular*}{\textwidth}{@{\extracolsep{\fill}}p{10pt}|p{35pt}p{35pt}p{35pt}|p{35pt}p{35pt}p{35pt}|p{35pt}p{35pt}p{35pt}|
p{35pt}p{35pt}p{35pt}} \hline\hline &
\multicolumn{3}{p{117pt}|}{\textit{T}=0~K} &
\multicolumn{3}{p{114pt}|}{\textit{T}=77~K} &
\multicolumn{3}{p{114pt}|}{\textit{T}=300~K} &
\multicolumn{3}{p{114pt}|}{\textit{T}=600~K}  \\ \hline
{\textit{n}}& \textit{S}$_{1/2}$& \textit{P}$_{1/2}$&
\textit{D}$_{3/2}$& \textit{S}$_{1/2}$& \textit{P}$_{1/2}$&
\textit{D}$_{3/2}$& \textit{S}$_{1/2}$& \textit{P}$_{1/2}$&
\textit{D}$_{3/2}$& \textit{S}$_{1/2}$& \textit{P}$_{1/2}$&
\textit{D}$_{3/2}$ \\ \cline{2-13}
 &
& \textit{P}$_{3/2}$& \textit{D}$_{5/2}$& & \textit{P}$_{3/2}$&
\textit{D}$_{5/2}$& & \textit{P}$_{3/2}$& \textit{D}$_{5/2}$& &
\textit{P}$_{3/2}$& \textit{D}$_{5/2}$ \\ \hline 10& 0.6260&
2.2016& 2.6317& 0.6247& 2.1792& 2.5882& 0.5854& 1.7812& 2.2115&
0.5217& 1.3424& 1.7692 \\ \hline --& --& 2.1431& 2.6011& --&
2.1218& 2.5585& --& 1.7435& 2.1900& --& 1.3217& 1.7556 \\ \hline
15& 2.7045& 8.7975& 7.6547& 2.6213& 7.9997& 7.2067& 2.2384&
5.4239& 5.5668& 1.8397& 3.6878& 4.1666 \\ \hline --& --& 8.5572&
7.5885& --& 7.8021& 7.1482& --& 5.3361& 5.5324& --& 3.6489& 4.1478
\\ \hline 20& 7.2216& 22.728& 17.556& 6.7623& 18.912& 15.767&
5.3824& 11.375& 11.177& 4.1743& 7.2867& 7.9154 \\ \hline --& --&
22.098& 17.414& --& 18.480& 15.654& --& 11.225& 11.121& --&
7.2271& 7.8881 \\ \hline 25& 14.902& 46.691& 33.732& 13.527&
35.877& 28.941& 10.175& 19.731& 19.096& 7.5607& 12.145& 12.959 \\
\hline --& --& 45.391& 33.469& --& 35.118& 28.748& --& 19.508&
19.014& --& 12.062& 12.922 \\ \hline 30& 27.005& 82.702& 57.935&
23.779& 59.263& 47.550& 16.965& 30.449& 29.468& 12.150& 18.233&
19.322 \\ \hline --& --& 80.397& 57.489& --& 58.094& 47.251& --&
30.146& 29.357& --& 18.126& 19.276 \\ \hline 35& 44.344& 134.72&
91.709& 37.931& 90.324& 72.139& 25.813& 43.752& 42.321& 17.931&
25.626& 26.990 \\ \hline --& --& 130.95& 91.008& --& 88.647&
71.708& --& 43.363& 42.178& --& 25.492& 26.935 \\ \hline 40&
67.858& 205.00& 136.72& 56.451& 129.20& 103.24& 36.813& 59.567&
57.695& 24.923& 34.289& 35.964 \\ \hline --& --& 199.25& 135.68&
--& 126.93& 102.65& --& 59.085& 57.519& --& 34.126& 35.899 \\
\hline 45& 98.502& 296.32& 194.40& 79.773& 176.25& 141.17& 50.040&
77.903& 75.580& 33.142& 44.224& 46.228 \\ \hline --& --& 287.99&
192.93& --& 173.32& 140.40& --& 77.324& 75.369& --& 44.031& 46.154
\\ \hline 50& 137.21& 411.18& 266.59& 108.27& 231.65& 186.42&
65.548& 98.753& 96.025& 42.598& 55.425& 57.795 \\ \hline --& --&
399.59& 264.58& --& 227.98& 185.45& --& 98.071& 95.779& --&
55.200& 57.713 \\ \hline 55& 184.93& 552.87& 354.26& 142.29&
295.81& 239.00& 83.379& 122.15& 118.96& 53.296& 67.905& 70.637 \\
\hline --& --& 537.28& 351.59& --& 291.35& 237.79& --& 121.37&
118.68& --& 67.646& 70.545 \\ \hline 60& 242.60& 723.70& 459.65&
182.15& 368.77& 299.46& 103.56& 148.08& 144.48& 65.241& 81.654&
84.779 \\ \hline --& --& 703.26& 456.19& --& 363.44& 298.02& --&
147.18& 144.16& --& 81.360& 84.679 \\ \hline 65& 311.16& 926.40&
584.42& 228.11& 450.67& 368.06& 126.14& 176.54& 172.57& 78.438&
96.674& 100.22 \\ \hline --& --& 900.22& 580.02& --& 444.42&
366.35& --& 175.53& 172.22& --& 96.343& 100.11 \\ \hline 70&
391.56& 1163.8& 730.09& 280.45& 541.65& 444.94& 151.11& 207.54&
203.25& 92.889& 112.97& 116.96 \\ \hline --& --& 1130.9& 724.61&
--& 534.43& 442.94& --& 206.41& 202.86& --& 112.60& 116.84 \\
\hline 75& 484.73& 1438.8& 898.07& 339.39& 641.87& 530.20& 178.51&
241.08& 236.47& 108.60& 130.53& 134.98 \\ \hline --& --& 1398.1&
891.34& --& 633.63& 527.89& --& 239.83& 236.05& --& 130.12& 134.86
\\ \hline 80& 591.62& 1754.0& 1090.1& 405.15& 751.35& 624.03&
208.35& 277.16& 272.27& 125.56& 149.37& 154.30 \\ \hline --& --&
1704.3& 1081.9& --& 742.05& 621.40& --& 275.78& 271.81& --&
148.92& 154.17 \\ \hline \hline
\end{tabular*}
\end{table*}

\bigskip

\begin{table*}
\caption{Effective lifetimes $\tau _{eff} \left( {\mu s} \right)$
of Rb \textit{nS}, \textit{nP}, and \textit{nD} Rydberg states.}
\begin{tabular*}{\textwidth}{@{\extracolsep{\fill}}p{10pt}|p{35pt}p{35pt}p{35pt}|p{35pt}p{35pt}p{35pt}|p{35pt}
p{35pt}p{35pt}| p{35pt}p{35pt}p{35pt}} \hline \hline &

\multicolumn{3}{p{117pt}|}{\textit{T}=0~K} &
\multicolumn{3}{p{114pt}|}{\textit{T}=77~K} &
\multicolumn{3}{p{114pt}|}{\textit{T}=300~K} &
\multicolumn{3}{p{114pt}}{\textit{T}=600~K}  \\ \hline
{\textit{n}}& \textit{S}$_{1/2}$& \textit{P}$_{1/2}$&
\textit{D}$_{3/2}$& \textit{S}$_{1/2}$& \textit{P}$_{1/2}$&
\textit{D}$_{3/2}$& \textit{S}$_{1/2}$& \textit{P}$_{1/2}$&
\textit{D}$_{3/2}$& \textit{S}$_{1/2}$& \textit{P}$_{1/2}$&
\textit{D}$_{3/2}$ \\ \cline{2-13}
 &
& \textit{P}$_{3/2}$& \textit{D}$_{5/2}$& & \textit{P}$_{3/2}$&
\textit{D}$_{5/2}$& & \textit{P}$_{3/2}$& \textit{D}$_{5/2}$& &
\textit{P}$_{3/2}$& \textit{D}$_{5/2}$ \\ \hline 10& 0.4639&
1.0768& 0.8264& 0.4637& 1.0741& 0.8230& 0.4427& 0.9862& 0.7797&
0.3992& 0.8388& 0.7118 \\ \hline --& --& 0.9994& 0.8201& --&
0.99702& 0.8167& --& 0.92246& 0.7744& --& 0.79419& 0.7076 \\ \hline
15& 2.3398& 4.7424& 2.8489& 2.2774& 4.5408& 2.7839& 1.9487&
3.6041& 2.4807& 1.6008& 2.7376& 2.1319 \\ \hline --& --& 4.4133&
2.8285& --& 4.2420& 2.7648& --& 3.4228& 2.4667& --& 2.6378& 2.1225
\\ \hline 20& 6.5768& 12.903& 6.9892& 6.1684& 11.693& 6.6755&
4.8990& 8.3288& 5.6284& 3.7905& 5.8832& 4.5986 \\ \hline --& --&
12.003& 6.9393& --& 10.965& 6.6316& --& 7.9749& 5.5999& --&
5.7165& 4.5816 \\ \hline 25& 14.332& 27.420& 13.995& 12.973&
23.534& 13.058& 9.6601& 15.352& 10.474& 7.1242& 10.296& 8.2014 \\
\hline --& --& 25.491& 13.895& --& 22.139& 12.975& --& 14.784&
10.425& --& 10.056& 8.1755 \\ \hline 30& 26.594& 50.010& 24.647&
23.283& 40.764& 22.471& 16.392& 24.730& 17.218& 11.635& 15.964&
13.004 \\ \hline --& --& 46.497& 24.471& --& 38.476& 22.331& --&
23.926& 17.144& --& 15.651& 12.968 \\ \hline 35& 44.386& 82.340&
39.679& 37.653& 63.935& 35.366& 25.223& 36.495& 25.988& 17.352&
22.880& 19.032 \\ \hline --& --& 76.549& 39.394& --& 60.510&
35.152& --& 35.431& 25.885& --& 22.491& 18.985 \\ \hline 40&
68.751& 126.54& 59.946& 56.588& 93.766& 52.256& 36.254& 50.745&
36.931& 24.298& 31.073& 26.328 \\ \hline --& --& 117.62& 59.516&
--& 88.957& 51.946& --& 49.403& 36.795& --& 30.606& 26.271 \\
\hline 45& 100.71& 184.36& 86.146& 80.529& 130.63& 73.485& 49.554&
67.478& 50.098& 32.484& 40.531& 34.891 \\ \hline --& --& 171.34&
85.526& --& 124.21& 73.058& --& 65.844& 49.926& --& 39.985& 34.823
\\ \hline 50& 141.31& 256.28& 126.53& 109.87& 174.26& 105.23&
65.176& 86.547& 68.939& 41.921& 51.199& 46.860 \\ \hline --& --&
239.23& 118.21& --& 166.53& 98.874& --& 84.742& 65.352& --&
50.618& 44.657 \\ \hline 55& 191.51& 346.17& 168.53& 144.94&
225.97& 137.30& 83.151& 108.23& 87.226& 52.609& 63.170& 58.251 \\
\hline --& --& 321.76& 158.32& --& 215.72& 129.73& --& 105.980&
83.124& --& 62.464& 55.779 \\ \hline 60& 252.44& 457.66& 218.98&
186.12& 286.63& 174.82& 103.53& 132.63& 107.92& 64.561& 76.478&
70.941 \\ \hline --& --& 422.88& 206.67& --& 273.09& 165.96& --&
129.84& 103.29& --& 75.611& 68.200 \\ \hline 65& 325.03& 587.99&
278.60& 233.62& 354.55& 218.06& 126.32& 159.34& 131.05& 77.774&
90.979& 84.929 \\ \hline --& --& 543.27& 263.98& --& 338.39&
207.83& --& 156.2& 125.89& --& 90.020& 81.919 \\ \hline 70&
410.41& 740.58& 348.19& 287.78& 430.64& 267.29& 151.55& 188.53&
156.63& 92.257& 106.73& 100.22 \\ \hline --& --& 684.24& 331.02&
--& 411.67& 255.61& --& 185.02& 150.93& --& 105.68& 96.940 \\
\hline 75& 509.57& 917.87& 428.45& 348.80& 515.28& 322.71& 179.25&
220.24& 184.69& 108.01& 123.75& 116.81 \\ \hline --& --& 848.00&
408.48& --& 493.32& 309.51& --& 216.37& 178.44& --& 122.60& 113.26
\\ \hline 80& 623.54& 1121.5& 500.91& 416.89& 608.55& 371.69&
209.42& 254.46& 208.93& 125.03& 142.02& 131.03 \\ \hline --& --&
1042.6& 497.28& --& 585.48& 369.83& --& 250.60& 208.45& --&
140.90& 130.89 \\ \hline \hline
\end{tabular*}
\end{table*}

\begin{table*}
\caption{Effective lifetimes $\tau _{eff} \left( {\mu s} \right)$
of Cs \textit{nS}, \textit{nP}, and \textit{nD} Rydberg states.}
\begin{tabular*}{\textwidth}{@{\extracolsep{\fill}}p{10pt}|p{35pt}p{35pt}p{35pt}|p{35pt}p{35pt}p{35pt}|p{35pt}p{35pt}p{35pt}|
p{35pt}p{35pt}p{35pt}}

\hline  \hline & \multicolumn{3}{p{117pt}|}{\textit{T}=0~K} &
\multicolumn{3}{p{114pt}|}{\textit{T}=77~K} &
\multicolumn{3}{p{114pt}|}{\textit{T}=300~K} &
\multicolumn{3}{p{114pt}}{\textit{T}=600~K}  \\ \hline
{\textit{n}}& \textit{S}$_{1/2}$& \textit{P}$_{1/2}$&
\textit{D}$_{3/2}$& \textit{S}$_{1/2}$& \textit{P}$_{1/2}$&
\textit{D}$_{3/2}$& \textit{S}$_{1/2}$& \textit{P}$_{1/2}$&
\textit{D}$_{3/2}$& \textit{S}$_{1/2}$& \textit{P}$_{1/2}$&
\textit{D}$_{3/2}$ \\ \cline{2-13}
 &
& \textit{P}$_{3/2}$& \textit{D}$_{5/2}$& & \textit{P}$_{3/2}$&
\textit{D}$_{5/2}$& & \textit{P}$_{3/2}$& \textit{D}$_{5/2}$& &
\textit{P}$_{3/2}$& \textit{D}$_{5/2}$ \\ \hline 10& 0.2932&
0.8936& 0.3094& 0.2932& 0.8930& 0.3092& 0.2869& 0.8611& 0.3031&
0.2667& 0.7658& 0.2909 \\ \hline --& --& 1.0165& 0.3176& --&
1.0155& 0.3173& --& 0.9720& 0.3109& --& 0.8525& 0.2981 \\ \hline
15& 1.7464& 4.6498& 1.3322& 1.7138& 4.5365& 1.3186& 1.5033&
3.7020& 1.2383& 1.2634& 2.8115& 1.1300 \\ \hline --& --& 5.1490&
1.3573& --& 5.0024& 1.3431& --& 4.0017& 1.2603& --& 2.9832& 1.1492
\\ \hline 20& 5.2626& 13.287& 3.5556& 4.9874& 12.308& 3.4701&
4.0594& 8.8255& 3.1316& 3.2040& 6.2018& 2.7392 \\ \hline --& --&
14.612& 3.6167& --& 13.407& 3.5281& --& 9.3628& 3.1804& --&
6.4560& 2.7791 \\ \hline 25& 11.912& 28.694& 7.4580& 10.913&
25.129& 7.1708& 8.3287& 16.400& 6.2374& 6.2529& 10.962& 5.2594 \\
\hline --& --& 31.329& 7.5803& --& 27.059& 7.2839& --& 17.160&
6.3267& --& 11.270& 5.3284 \\ \hline 30& 22.636& 53.372& 13.532&
20.086& 44.182& 12.816& 14.492& 26.684& 10.771& 10.454& 17.158&
8.7940 \\ \hline --& --& 58.592& 13.746& --& 47.568& 13.009& --&
27.790& 10.914& --& 17.552& 8.8992 \\ \hline 35& 38.399& 90.019&
22.260& 33.056& 70.494& 20.771& 22.691& 39.801& 16.901& 15.842&
24.804& 13.412 \\ \hline --& --& 97.999& 22.607& --& 75.065&
21.074& --& 41.048& 17.114& --& 25.186& 13.560 \\ \hline 40&
60.186& 139.52& 34.107& 50.330& 103.83& 31.364& 33.032& 55.491&
24.757& 22.442& 33.779& 19.157 \\ \hline --& --& 151.80& 34.618&
--& 110.10& 31.799& --& 56.961& 25.047& --& 34.173& 19.351 \\
\hline 45& 88.953& 204.59& 49.562& 72.339& 145.03& 44.925& 45.590&
73.906& 34.462& 30.266& 44.136& 26.071 \\ \hline --& --& 222.68&
50.297& --& 153.33& 45.533& --& 75.601& 34.848& --& 44.531& 26.319
\\ \hline 50& 125.64& 287.56& 69.099& 99.463& 194.60& 61.756&
60.414& 95.070& 46.112& 39.321& 55.876& 34.182 \\ \hline --& --&
312.90& 70.116& --& 205.06& 62.575& --& 96.944& 46.606& --&
56.248& 34.489 \\ \hline 55& 171.30& 390.31& 93.227& 132.11&
252.69& 82.169& 77.569& 118.95& 59.801& 49.624& 68.982& 43.520 \\
\hline --& --& 424.65& 94.593& --& 265.50& 83.240& --& 120.98&
60.417& --& 69.315& 43.891 \\ \hline 60& 226.86& 514.90& 122.40&
170.58& 319.55& 106.42& 97.082& 145.53& 75.589& 61.174& 83.450&
54.096 \\ \hline --& --& 560.22& 124.19& --& 334.90& 107.78& --&
147.70& 76.340& --& 83.730& 54.535 \\ \hline 65& 293.29& 663.62&
157.11& 215.16& 395.46& 134.77& 118.98& 174.84& 93.540& 73.976&
99.281& 65.926 \\ \hline --& --& 722.23& 159.40& --& 413.53&
136.47& --& 177.13& 94.435& --& 99.495& 66.436 \\ \hline 70&
371.58& 838.67& 197.83& 266.13& 480.61& 167.46& 143.28& 206.88&
113.71& 88.034& 116.48& 79.023 \\ \hline --& --& 912.81& 200.70&
--& 501.50& 169.54& --& 209.24& 114.76& --& 116.61& 79.606 \\
\hline 75& 462.69& 1042.0& 245.06& 323.71& 575.10& 204.75& 170.01&
241.62& 136.14& 103.35& 135.03& 93.398 \\ \hline --& --& 1134.1&
248.60& --& 598.87& 207.25& --& 244.03& 137.36& --& 135.06& 94.058
\\ \hline 80& 567.60& 1275.9& 299.26& 388.13& 679.12& 246.83&
199.18& 279.08& 160.87& 119.93& 154.94& 109.06 \\ \hline --& --&
1388.3& 303.57& --& 705.74& 249.80& --& 281.50& 162.26& --&
154.85& 109.80 \\ \hline \hline
\end{tabular*}
\end{table*}

\end{document}